\documentclass[twocolumn,superscriptaddress,aps,preprintnumbers,amsmath,amssymb,sort&compress,nofootinbib,prl]{revtex4-2}
 
\usepackage{amsfonts,amsmath,amssymb,ascmac,bm,tensor}
\usepackage{fnpct} 
\usepackage{comment}
\usepackage{ifpdf}
\usepackage{slashed}
\usepackage{color}
\usepackage[mathscr]{eucal}
\usepackage[utf8]{inputenc}
\usepackage{physics}
\usepackage{cancel}

\ifpdf
  \usepackage{graphicx}     %   usepackage without driver option
  \usepackage[bookmarksopen,colorlinks=true,linkcolor=bblue,citecolor=bblue,urlcolor=ppink]{hyperref}
\else     % For (p)LaTeX + dvipdfmx
\fi

\definecolor{red}{rgb}{1,0,0}
\definecolor{darkred}{rgb}{0.6,0,0}
\definecolor{darkgreen}{rgb}{0.992447,0.623778,0.034597}
\definecolor{ppink}{rgb}{1,0.4,0.4}
\definecolor{bblue}{rgb}{0.284602,0.317763,0.963947}
\definecolor{purple}{rgb}{0.5 ,0, 0.7}

\newcommand{\ee}{\text{e}}
\newcommand{\Pl}{\text{Pl}}
\newcommand{\CMB}{\text{CMB}}

\makeatletter
\newcommand\footnoteref[1]{\protected@xdef\@thefnmark{\ref{#1}}\@footnotemark}
\makeatother

\allowdisplaybreaks[1]

\begin{document}

%%%%%%%%%%%%%%%%%%%%%%%%%%%
%%%%%%%%%%% Title %%%%%%%%%%%
%%%%%%%%%%%%%%%%%%%%%%%%%%%

\title{
Primordial Black Holes Arise When The Inflaton Falls
}

\author{Keisuke Inomata}
\affiliation{Kavli Institute for Cosmological Physics and Enrico Fermi Institute, The University of Chicago, Chicago, IL 60637, USA}
\author{Evan McDonough}
\affiliation{Kavli Institute for Cosmological Physics and Enrico Fermi Institute, The University of Chicago, Chicago, IL 60637, USA}
\author{Wayne Hu}
\affiliation{Kavli Institute for Cosmological Physics and Enrico Fermi Institute, The University of Chicago, Chicago, IL 60637, USA}
\affiliation{Department of Astronomy \& Astrophysics, The University of Chicago, Chicago, IL 60637, USA}

\begin{abstract}
\noindent Primordial Black Holes (PBHs) have entered the forefront of theoretical cosmology, due their potential role in phenomena ranging from gravitational waves, to dark matter, to galaxy formation. 
While producing PBHs from inflationary fluctuations naively would seem to require a large deceleration of the inflaton from its velocity at the horizon exit of CMB scales, in this work we demonstrate that an acceleration from a relatively small downward step in the potential that is transited in much less than an e-fold amplifies fluctuations as well. Depending on the location of the step, such PBHs could explain dark matter or the black holes detected by the gravitational wave interferometers. The perturbation enhancement has a natural interpretation as particle production due to the non-adiabatic transition associated with the step.
\end{abstract}

\date{\today}
\maketitle

\emph{Introduction.}---
Primordial black holes (PBHs) are one of the most intriguing topics in modern cosmology, owing to their potential to explain dark matter (DM) and the BHs detected by the LIGO-Virgo collaboration~\cite{Bird:2016dcv,Clesse:2016vqa,Sasaki:2016jop,Garcia-Bellido:2020pwq}.
Also, PBHs might be related to other observational results, such as the existence of supermassive black holes~\cite{Duechting:2004dk,Kawasaki:2012kn,Nakama:2016kfq,Hasegawa:2017jtk,Kawasaki:2019iis,Kitajima:2020kig,Shinohara:2021psq}, the OGLE results~\cite{mroz2017no,Niikura:2019kqi}, the recent NANOGrav results~\cite{Arzoumanian:2020vkk,Vaskonen:2020lbd,DeLuca:2020agl,Kohri:2020qqd,Sugiyama:2020roc,Domenech:2020ers,Inomata:2020xad,Atal:2020yic,Kawasaki:2021ycf}, and the anomalous excess of 511\,keV photons~\cite{Keith:2021guq} (see also Refs.~\cite{Sasaki:2018dmp,Carr:2020gox,Green:2020jor} for recent reviews). 
PBHs can be produced when very large density perturbations enter the horizon in the early universe.
In particular, the PBH scenarios for DM or LIGO-Virgo events can be associated with the large power spectrum of primordial curvature perturbations, $\mathcal P_{\mathcal R} \sim 10^{-2}$~\cite{Sasaki:2018dmp}, on small scales.

Throughout this letter, we focus on single-field inflation models that can realize the large power spectrum on small scales for the PBH scenarios. Under the slow-roll approximation, the power spectrum  is given by $\mathcal P_{\mathcal R} = H^2_*/(8\pi^2 M_\Pl^2 \epsilon_*)$, where the subscript ``$*$'' denotes evaluation at the horizon exit of the perturbation and $\epsilon\equiv  -\dd \ln H /\dd N = (\dd \phi/\dd N)^2/(2M_\Pl^2)$, where $N\equiv \int H {\rm d}t$ is the number of e-folds of inflationary expansion.
From this relation, at first glance, the large power spectrum on small scales needed for the PBH scenarios seems to require a substantial decrease in $\epsilon$, and hence the kinetic energy of the inflaton, from the horizon exit of CMB scales.  
This decrease is realized by a large negative value of $\eta \equiv \dd \ln \epsilon/\dd N$ which violates the slow-roll assumption~\cite{Motohashi:2017kbs}. 
This can be achieved with a very flat potential in a period of so called ``ultra slow roll (USR)'' when Hubble friction dominates over the potential slope~\cite{Ivanov:1994pa,Inoue:2001zt,Tsamis:2003px,Kinney:2005vj,Garcia-Bellido:2017mdw,Ezquiaga:2017fvi,Kannike:2017bxn,Germani:2017bcs,Ballesteros:2017fsr,Hertzberg:2017dkh,Byrnes:2018txb,Passaglia:2018ixg}.  
On the other hand since the slow-roll approximation must be violated~\cite{Motohashi:2017kbs}, this invalidates the naive expectation of a decreased $\epsilon$ and leaves the possibility of alternative mechanisms.

\begin{figure} [ht!]
\centering \includegraphics[width=\columnwidth]{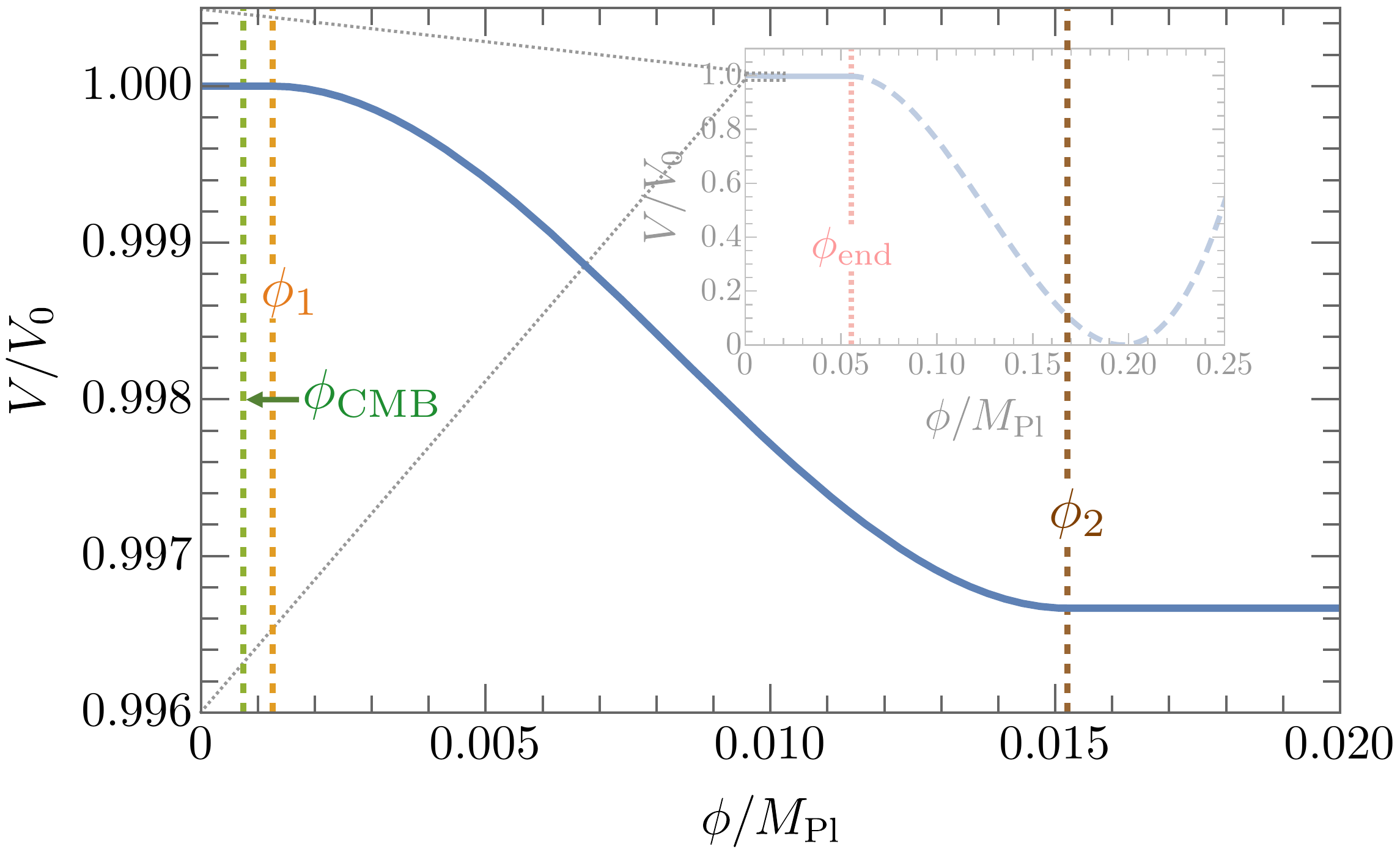}
\caption{The inflaton potential of Eq.~(\ref{eq:pot_cmb_to_end_mo}) 
that realizes the large enhancement of perturbations, with   the steplike transition at $\phi_1 \le \phi \le \phi_2$ highlighted and an inset for the full range.
The parameters are $n_s = 0.97$, $\epsilon_1=7.43\times 10^{-10}$, $\epsilon_2 = 0.01$, $\epsilon_3 = 10^{-9}$, and $\Delta N_\text{step} = 0.5$. $\phi_\text{end}$ denotes the end of inflation (red vertical dotted line) and corresponds to $50$ e-folds from $\phi_\text{CMB}$. 
}
\vspace{-0.5cm}
\label{fig:pot_l01}
\end{figure}

In this letter, we show that a decrease in the kinetic energy of the inflaton relative to that at CMB scales is not necessary  for the large enhancement of perturbations required for the PBH scenarios. 
Equivalently, the inflation potential need not have a region that is flatter than it is at CMB scales. 
If the inflaton instead {\it gains} kinetic energy by rolling down a sufficiently sharp feature that it crosses in less than an e-fold, non-adiabatic particle production occurs.  
Furthermore if there is thereafter a period of slow-roll inflation where $\epsilon$ returns to near its previous value,  this particle production produces an oscillatory enhancement of the power spectrum whose size reflects the ratio of kinetic energies before and after the transition \cite{Miranda:2015cea}.
Fig.~\ref{fig:pot_l01} shows one example of such an inflaton potential that realizes $\mathcal P_\mathcal R \sim 10^{-2}$ on small scales consistently with the CMB measurements.
The specific form of the potential is given later in Eq.~(\ref{eq:pot_cmb_to_end_mo}).

\emph{Mechanism.}---
In canonical single-field inflation, the Fourier mode of the comoving curvature perturbation ${\cal R}_k$ obeys the Mukhanov-Sasaki equation~\cite{Sasaki:1986hm,Mukhanov:1988jd}:
\begin{align}
	\mathcal R_k'' + \left(2 + \eta \right) aH \mathcal R_k' + k^2 \mathcal R_k = 0,
	\label{eq:r_curv_eom_com}
\end{align}
where $k$ is the comoving wavenumber,  $a$ is the scale factor, and the prime denotes the derivative with respect to the conformal time, ${\rm d} \tau\equiv {\rm d}t/a$ with $\tau=0$ at the end of inflation. The Bunch-Davies vacuum  provides the initial condition $\mathcal R_k(\tau_\text{ini}) = \frac{H_\text{ini}}{\sqrt{4k^3 \epsilon_\text{ini}} M_\Pl} (-k \tau_\text{ini}) \ee^{-i k\tau_\text{ini}}$, where $k |\tau_\text{ini}| \gg 1$ and the subscript ``ini'' indicates the value at the initial time.

The dynamics of ${\cal R}_k$ are completely determined by the expansion history, as encoded in Eq.~\eqref{eq:r_curv_eom_com} by $\eta$, $a$, and $H$.
In this work we develop a mechanism for perturbation enhancement that can be concisely specified by a simple background evolution:
at the end of an  initial stage of inflation the slow-roll parameter $\epsilon$ reaches $\epsilon_1$, as inherited from CMB scales. At this time, $\epsilon$ rapidly increases to a value of $\epsilon_2$ in much less than an e-fold so that $\eta \gg 1$.  Following this, the universe enters a USR phase, wherein $\eta=-6$~\cite{Kinney:2005vj,Byrnes:2018txb}, and $\epsilon$ decreases from $\epsilon_2$ back to a value  $\epsilon_3
  \sim \epsilon_1$. 

This background evolution can be explicitly realized by an inflaton potential that exhibits a downward step, as shown in Fig.~\ref{fig:pot_l01}. Inflation with this potential generates a spike in the primordial power spectrum, and a pattern of oscillations on small scales. In what follows we will demonstrate this both analytically for a toy model, and numerically for an explicit choice of the inflaton potential.

{\it Toy Model with an Analytic Solution.}---
To understand the mechanism analytically, we can approximate the evolution of ${\cal R}_k$ by a piecewise sequence of periods where $\eta\approx\,$const.\ while $\epsilon \ll 1$.  Under these conditions the general solution for Eq.~\eqref{eq:r_curv_eom_com} is
\begin{align}
	\mathcal R_k \approx C_1 G^{(1)}_\nu(-k \tau) + C_2 G^{(2)}_\nu(-k \tau),
	\label{eq:curv_r_general}
\end{align}
where $\nu = 3/2 + \eta/2$, $C_1$ and $C_2$ are constant in time, and $G^{(1)}_\nu$ and $G^{(2)}_\nu$ are defined with the Hankel functions of the first ($H_\nu^{(1)}$) and the second kind ($H_\nu^{(2)}$) as
\begin{align}
	G^{(j)}_\nu(-k \tau) \equiv  (-k\tau)^\nu H_\nu^{(j)}(-k \tau),
\end{align}
with $j \in (1,2)$.

We consider the case where $\epsilon$ changes from $\epsilon_1 (\ll 1)$ to $\epsilon_2 (>\epsilon_1)$ with constant positive $\eta$ and, after that, $\epsilon$ decreases with $\eta = -6$, which corresponds to a USR phase. Specifically, we parameterize $\eta$ as 
\begin{align}
	\eta = \eta_c \Theta(\tau-\tau_1) \Theta(\tau_2 -\tau) -6 \Theta(\tau -\tau_2).
	\label{eq:eta_two_periods}
\end{align}
Once $\epsilon_1/\epsilon_2$ is fixed, we have $\tau_2/\tau_1 = (\epsilon_1/\epsilon_2)^{1/\eta_c}$ and $\tau/\tau_2 = (\epsilon(\tau)/\epsilon_2)^{1/6}$ for $\tau>\tau_2$ with $\epsilon(\tau)$ being the value at $\tau$. Then, the solution of the curvature perturbation is given by 
\begin{align} 	
\mathcal R_k = 
\begin{cases}
	D_1 G^{(1)}_{3/2}(-k \tau) & (\tau < \tau_1)\\	
	E_1 G^{(1)}_{\nu_c}(-k \tau)  + E_2 G^{(2)}_{\nu_c}(-k \tau)  & (\tau_1 \leq \tau \leq \tau_2)\\
	F_1 G^{(1)}_{-3/2}(-k \tau)  + F_2 G^{(2)}_{-3/2}(-k \tau)  & (\tau_2 \leq \tau)  
	\end{cases},  \nonumber
\end{align}
where $\nu_c = 3/2 + \eta_c/2$. The coefficient $D_1$ is determined by the Bunch-Davies vacuum condition as $D_1 = - \sqrt{{\pi}/{2}} H/(\sqrt{4k^3 \epsilon_1}M_\Pl)$, while the other coefficients are determined by matching conditions, namely the continuity of $\mathcal R$ and $\mathcal R'$ at $\tau_1$ and $\tau_2$. These matching conditions are the linearized Israel junction conditions \cite{1966NCimB..44....1I,Deruelle:1995kd}  for the metric across the constant-$\tau$ hypersurfaces $\tau=\tau_1$, $\tau_2$.

The dimensionless power spectrum is given by
\begin{align}
	\mathcal P_\mathcal R(k) \equiv&  \frac{k^3}{2\pi^2} |\mathcal R_k(\tau_3)|^2,
	\label{eq:p_r_analytical_3}
\end{align}
where $\mathcal R_k$ is evaluated at an epoch $\tau_3$ after which the curvature has frozen out.   In our toy model we take this to be at the end of the USR phase so that
$\epsilon_3 = \epsilon(\tau_3)$.   We shall see that for a step potential there is a second slow roll phase once $\epsilon(\tau) \approx \epsilon_1$.

This toy model admits an analytic solution in the limit that $\epsilon$ undergoes an instantaneous transition from $\epsilon_1$ to $\epsilon_2$. This corresponds to the limit $|\tau_1 - \tau_2 | \rightarrow 0$, $\eta_c \rightarrow \infty$, with $\eta_c |\tau_1 -\tau_2|$ held fixed. 
In this limit, Eq.~(\ref{eq:r_curv_eom_com}) becomes 
\begin{equation}
\mathcal{R}_k'' + \frac{\epsilon'}{\epsilon} \mathcal R_k'\approx 0, 
\end{equation}
for $\tau_1 <\tau < \tau_2$, up to corrections that scale as $k^2/(\eta_c aH)^2 $. 
From its solution  $ \epsilon(\tau) {\cal R}_k ' (\tau) = {\rm constant} $, one may deduce the relation between $\epsilon$ and ${\cal R}$ on either side of the transition, as ${\mathcal R}_k'(\tau_2) = ({\epsilon_1}/{\epsilon_2}) {\mathcal R}_k'(\tau_1)$. Thus, ${\cal R}_k'$ undergoes a jump down at the transition, by a relative factor of $(\epsilon_1/ \epsilon_2)$, while ${\cal R}$ itself is continuous, ${\mathcal R}_k(\tau_1)= {\mathcal R}_k(\tau_2)$, for any finite $\epsilon$. 

In the limit $\epsilon_1/\epsilon_2 \rightarrow 0$, the ${\cal R}'$ after the transition vanishes, ${\cal R}'\rightarrow0$.   For $\tau>\tau_2$, the modes then behave like they began at $\tau_2$ in an excited state with an amplitude ${\mathcal R}_k \propto \epsilon_1^{-1/2}$ which is much higher  than the adiabatic prediction where ${\mathcal R}_k \propto \epsilon_2^{-1/2}$.
Moreover, matching the perturbations across the transition converts the incoming ($\tau<\tau_1$) positive frequency mode $G_{3/2} ^{(1)}$ into outgoing ($\tau > \tau_2 $) positive and negative frequency modes, $G_{-3/2} ^{(1)}$ and $G_{-3/2} ^{(2)}$. One may interpret this as particle production due to the adiabaticity violation at the transition \cite{Miranda:2015cea}.
Indeed, in the limit $\eta_c \rightarrow \infty$, one may explicitly solve for the coefficients of the positive and negative frequency modes of the outgoing state, $F_1$ and $F_2$, to find,
\begin{align}
	F_{1,\text{lim}} =& D_1 \frac{ 3\epsilon_2 + (2\epsilon_2 + \epsilon_1) (-k\tau_1)^2 + i(\epsilon_2 + \epsilon_1) (-k\tau_1)^3 }{2\epsilon_2}, \nonumber \\
	F_{2,\text{lim}} =& -D_1 \ee^{-2ik\tau_1} (1-i(-k\tau_1)) \nonumber \\
	& \  \times \frac{(3 \epsilon_2 - 3 i \epsilon_2(-k\tau_1) + (\epsilon_1 - \epsilon_2)(-k\tau_1)^2 ) }{2\epsilon_2},
\end{align}
where note again $\tau_2 = \tau_1$ in this limit.

In the large-scale limit $k\ll 1/|\tau_1|$, the power spectrum becomes the conventional expression for slow-roll inflation in a potential with slow-roll parameter $\epsilon_1$, up to corrections of order $\mathcal{O}( \sqrt{\epsilon_2 / \epsilon_3} (-k\tau_1)^2)$. We find,
\begin{align}
	\mathcal P_{\mathcal R}(k) \simeq & \frac{H^2}{8\pi^2 M_\Pl^2 \epsilon_1 } \left[ 1 - \frac{2}{15} \sqrt{\frac{\epsilon_2}{\epsilon_3}} (-k\tau_1)^2  \right. \nonumber \\
	& \left.  \qquad \quad \qquad
	+ \frac{1}{225} \frac{\epsilon_2}{\epsilon_3}(-k\tau_1)^4 \right],
	\label{eq:pr_large_limit_downward}	
\end{align}
up to $\mathcal O((-k\tau_1)^4)$, and where we have also assumed $\epsilon_3/\epsilon_2 \ll 1$. On the other hand, for modes which were inside the horizon at the transition ($k \gtrsim 1/|\tau_1|$), 
the power spectrum becomes
(see also Ref.~\cite{Miranda:2015cea})
\begin{align}
	\mathcal P_{\mathcal R}(k) \simeq 
	A \frac{1 - \cos(-2k\tau_1)}{2}, \quad A=\frac{H^2}{8\pi^2 M_\Pl^2 \epsilon_1 } \frac{\epsilon_2}{\epsilon_3},  
	\label{eq:pr_small_limit_downward}
\end{align}
where we have assumed $\epsilon_1/\epsilon_2 \ll 1$ and we have used the relation $(\tau_1/\tau_3)^6 \approx \epsilon_2/\epsilon_3$.
Therefore there is an $\mathcal O(\epsilon_2/\epsilon_3)$ enhancement of the power spectrum, and oscillations of frequency $\sim 1/(2|\tau_1|)$ in $k$.

Beyond this instantaneous transition limit, there are  corrections of order $(k\tau_1/\eta_c)^2$ to the equation of motion. 
Modes that oscillate much faster than the transition timescale evolve adiabatically and do not experience enhancement.   
In the intermediary regime where the modes oscillate a few times during the transition, the enhancement is damped  \cite{Adshead:2011jq}. Indeed, 
one can show that~\cite{Inomata:preparation}
\begin{equation}
\label{eq:damping}
    A \rightarrow |D(k,\tau_1, \eta_c)|^2 A,
\end{equation}
where
\begin{align}
		\label{eq:damping_fac}
	D(k,\tau_1, \eta_c) 
	&\simeq \left( \frac{\epsilon_2}{\epsilon_1} \right)^{ \sqrt{-\frac{(-k\tau_1)^2}{\eta_c^2} + \frac{3}{2\eta_c} + \frac{1}{4}} - \frac{1}{2}}.
\end{align}

For $\eta_c \gg 1$, as $|k \tau_1| \ll \eta_c$ , $|D|^2 \rightarrow 1$,
whereas, 
in the limit $| k\tau_1| \gg \eta_c$, 
 $|D|^2 \rightarrow \epsilon_1/\epsilon_2$  and removes the enhancement in Eq.~(\ref{eq:pr_small_limit_downward}).
Beyond the piecewise toy model constructed here where $\eta_c$ is constant, it is useful to relate the damping with an average
$\bar\eta$ that corresponds to the number of e-folds $\Delta{N_{\rm step}}$ for the inflaton to complete the transition
\begin{equation}
\label{eq:Nstepeta}
\eta_c \rightarrow \bar\eta \equiv \ln(\epsilon_2/\epsilon_1)/
\Delta N_{\rm step},
\end{equation}
which can be used to reparameterize Eq.~(\ref{eq:damping_fac}) with $\Delta N_\text{step}$.
We shall next see that these analytic estimates provide an excellent fit to the power spectrum in a concrete model.

\emph{A concrete model.}---
We start with a base potential that can satisfy CMB constraints as well as
end inflation at some $\phi_{\rm end}$.
For definiteness we take
\begin{align}
  V_b(\phi) \simeq V_0 \left(1 - \frac{\beta \phi^2/ M_\Pl^2}{1+\phi/\phi_\text{CMB}} \right),
  \label{eq:vbase_pot}
\end{align}
for $\phi \ll \phi_{\rm end}$. 
Here $\phi_\text{CMB}$ is the inflaton value at the horizon exit of the CMB normalization scale ($k=0.05\,$Mpc$^{-1}$) during the inflation.   Though this specific $\phi_{\rm CMB}$ dependent form is not required, this asymptotically linear  potential after $\phi_\text{CMB}$ simplifies  the relationship to $\epsilon$ and the comparison to the analytic results.
Near $\phi_{\rm end}$, $V_b$ changes form to some $V_{\rm end}$ in order to end inflation. 
Our specific choice is displayed in Fig.~\ref{fig:pot_l01}, but the detailed form is not important for our analysis, as it does not play a role in the enhancement of perturbations.

The enhancement of perturbations over the base model comes from a continuous but step-like transition
\begin{align}
	V(\phi) = V_b(\phi) F\left(\phi; \phi_1,\phi_2,
	h \right).
	 \label{eq:pot_cmb_to_end_mo}
\end{align}
In order to mimic the toy model where $\eta$ is nearly constant
during the transition, we take
\begin{equation}
	F(\phi;\phi_1,\phi_2,h) \equiv 
\begin{cases}
	1 & (\phi < \phi_1) \\
	1 - h S\left( \frac{\phi - \phi_1}{\phi_2-\phi_1}  \right) & 
	(\phi_1 \le \phi \le \phi_2 )  \\
	 1 - h &  (\phi> \phi_2 ) 
\end{cases},
\label{eq:all_potential}
\end{equation}
where $S(x) \equiv x^2(3- 2x)$ and it changes from $0$ to $1$ with the change of $x$ from $0$ to $1$.
The parameter $h$ corresponds to the height of the potential step normalized by $V_b$, whereas $\phi_1$ and $\phi_2$ its beginning and end.
Note that with Eq.~(\ref{eq:all_potential}), $V''$ is formally discontinuous at $\phi_1$ and $\phi_2$ though we have verified that a smooth $V''$ gives indistinguishable results as long as its net change occurs in much less than an e-fold. We will discuss this issue again later.

The base potential $V_b$ determines the tilt of power spectrum at the CMB scale and the slow-roll parameter $\epsilon$ before and long after the transition.
The tilt of power spectrum at the CMB scale ($n_s$) is mainly determined by the second derivative of the potential as 
\begin{align}
    n_s -1 \simeq 2 M_\Pl^2 \frac{V''(\phi_\text{CMB})}{V(\phi_\CMB)} \simeq -\frac{\beta}{2}.
\end{align}
We take the value of $\beta$ to be consistent with the CMB measurement.
The value of $\epsilon$ during the slow-roll phases (before and well after the transition) can be expressed as 
\begin{align}
    \epsilon \simeq \frac{M_\Pl^2}{2} \left( \frac{V'}{V} \right)^2 \simeq \frac{\beta^2\phi_\CMB^2}{2 M_\Pl^2} \left[\frac{ (\phi/\phi_\CMB) (2 + \phi/\phi_\text{CMB})}{(1 + \phi/\phi_\text{CMB} )^2} \right]^2.
\end{align}
We take $V_0/M_\Pl^4 = 24\pi^2 \epsilon(\phi_\text{CMB}) \times 2.1 \times 10^{-9}$ to be consistent with the CMB measurements~\cite{Aghanim:2018eyx}.

The potential can now be reparameterized using $\epsilon$.  
First, at the start of the transition $\epsilon_1 \equiv \epsilon(\phi_1)$; next well after the transition
$\epsilon_3= \beta^2\phi_\CMB^2/(2M_\Pl^2)$; finally right after the transition in the limit of $h\ll 1$ and 
$\Delta N_{\rm step} \ll 1$, energy conservation requires that 
\begin{equation}
    h=
	\frac{\epsilon_2 -\epsilon_1}{3}.
\end{equation}
We can also relate the step width $\phi_2-\phi_1$ to
$\Delta N_\text{step}$ through Eq.~(\ref{eq:Nstepeta}) with~\cite{Inomata:preparation}
\begin{equation}
    \bar \eta =   -6 + \frac{24hM_\Pl^2}{(\phi_2 - \phi_1)^2} \left( -1 + \sqrt{1+ \frac{24 hM_\Pl^2}{3 (\phi_2 - \phi_1)^2}} \right)^{-1}. 
	\label{eq:eta_app_exp_ease}
\end{equation}
The potential model is now parameterized by
the observed $n_s$, $\epsilon_1$, $\epsilon_2$, $\epsilon_3$, and $\Delta N_{\rm step}$.  Notice also that
$\epsilon_1 \sim \epsilon_3$, as long as 
$\phi_1 > \phi_\CMB$, which fulfills the requirements of the mechanism along with 
$\Delta N_{\rm step}\ll 1$ and $\epsilon_2/\epsilon_1\gg 1$.

\begin{figure} 
\centering \includegraphics[width=\columnwidth]{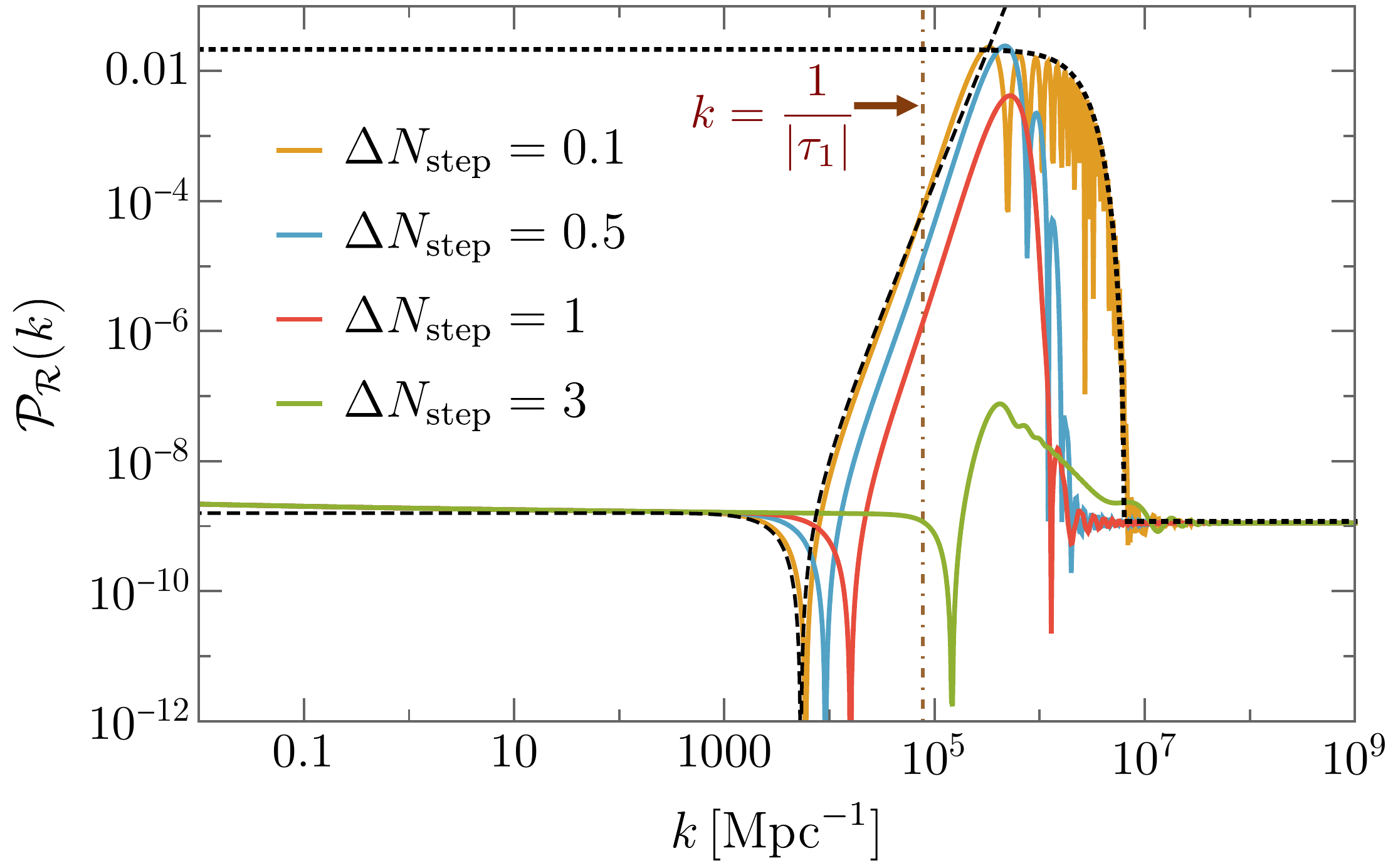}
\caption{ Curvature power spectrum for the potential of Eq.~(\ref{eq:pot_cmb_to_end_mo}) with different $\Delta N_\text{step}$ with the same parameters as Fig.~\ref{fig:pot_l01} otherwise.
For comparison, we also plot approximate formulas:
Eq.~(\ref{eq:pr_large_limit_downward}) 
on  large scales (black dashed) and 
Eq.~(\ref{eq:damping}) on small scales (black dotted,
with $\Delta N_\text{step}=0.1$)
 respectively.
 }
\label{fig:ps_cmb_tra}
\end{figure}

\begin{figure} 
\centering \includegraphics[width=\columnwidth]{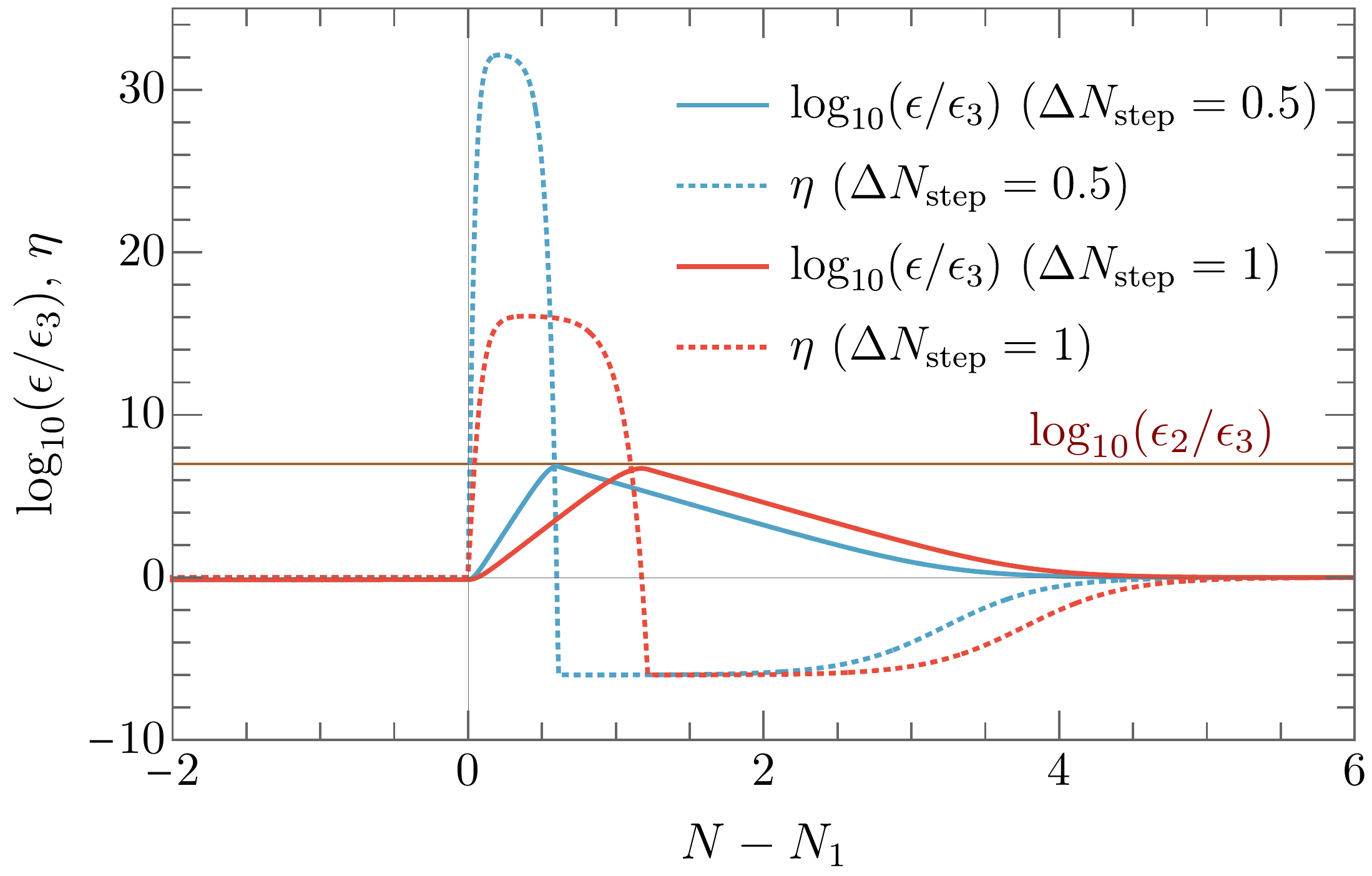}
\caption{ The evolution of $\epsilon$ and $\eta$ with different $\Delta N_\text{step}$ in e-folds from
$N_1=N(\tau_1)$. Except for $\Delta N_\text{step}$, we take the same parameters as in Fig.~\ref{fig:pot_l01}.
}
\label{fig:ep_eta_cmb_tra}
\end{figure}

Figure~\ref{fig:pot_l01} shows the potential that realizes the perturbation enhancement consistently with the CMB measurements.
Notice that although $\Delta N_{\rm step}=0.5$ and $\epsilon_2/\epsilon_1 \simeq 1.35 \times 10^7$ the transition in field space is neither sharp nor large: $\phi_2/\phi_1 \simeq 12.1$, $h\simeq 3.33\times 10^{-3}$.  This is because the kinetic energy suddenly increases from a very small initial value.
Figure~\ref{fig:ps_cmb_tra} shows the power spectra in the potential with different $\Delta N_\text{step}$.
The peak height of the power spectrum is mainly controlled by $\epsilon_2/\epsilon_3$ (or $h$) and the peak scale is determined by  $\epsilon_1$ (or
$\phi_1$) which then determines the PBH mass scale.
As a fiducial example,
we take parameters that realize $\mathcal P_{\mathcal R}(k\sim \mathcal O(10^5)\,\text{Mpc}^{-1}) \simeq \mathcal O(10^{-2})$, 
which corresponds to the LIGO-Virgo PBH scenario~\cite{Inomata:2017vxo,Sasaki:2018dmp}.\footnote{
This estimate does not take into account the effect of the non-Gaussianity~\cite{Byrnes:2012yx,Young:2013oia,Pattison:2017mbe,Biagetti:2018pjj,Ezquiaga:2018gbw,Ezquiaga:2019ftu,Figueroa:2020jkf,Pattison:2021oen}, which could change the specific amplitude required for a given PBH abundance.}
We can see that the approximate formulas, given in Eqs.~(\ref{eq:pr_large_limit_downward}) and (\ref{eq:damping}), fit the numerical result very well.
Figure~\ref{fig:ep_eta_cmb_tra} shows the evolution of $\epsilon$ and $\eta$, which indicates that  Eq.~(\ref{eq:eta_app_exp_ease}) is a good approximation to the steplike enhancement of $\eta$ whereas immediately after the step
$\eta \simeq -6$ for all the cases.

Although our  model provides a concrete realization of the enhancement, the mechanism itself is generic.   
The   power spectrum enhancement applies to  any base potential $V_b(\phi)$ where $\epsilon_1$ is sufficiently small so that $\epsilon_2 \ll 1$.
Our specific example shows that CMB constraints can also be accommodated.
As a  consequence, for PBHs to form abundantly $\epsilon_1$ must be so small as to make tensor fluctuations at CMB scales practically unobservable, which is a falsifiable prediction that distinguishes this mechanism from most PBH alternatives.
Furthermore, since the enhancement over CMB scales is by $\epsilon_2/\epsilon_3$, 
the mechanism can even accommodate a hierarchy  between $\epsilon_1$ and $\epsilon_3$ as long as $\epsilon_2 \ll 1$.

The form of
Eq.~(\ref{eq:all_potential}) actually encodes two types of transitions: one for $\epsilon$ occurring over $\Delta N_{\rm step}\ll 1$ and the other for
$V''$ (or relatedly $\eta$) at $\phi_{1,2}$ which we have for simplicity taken to occur over $\Delta N_{1,2}\rightarrow 0$. 
While this limit does not present a problem for solving Eq.~(\ref{eq:r_curv_eom_com}), it does present a strong-coupling problem due to nonlinearity in the full field equation.   Perturbation modes become strongly coupled when $V''(\phi=\bar\phi +\delta\phi)$ can no longer be approximated by the background field
$V''(\bar\phi)$. As shown in Ref.~\cite{Adshead:2014sga} this will occur
for modes of interest for transitions of width $\Delta N \lesssim 10^{-2}-10^{-3}$.   We have explicitly checked that smoothing these transitions to $\Delta N_1 \sim 10^{-1}$ and 
$\Delta N_2 \sim 10^{-2}$ yields nearly indistinguishable enhancements (see Ref.~\cite{Inomata:preparation}).   
Similarly, though the transitions in $\epsilon$ and $\eta$ may be combined in other models  (see Ref.~\cite{Miranda:2015cea})  into a single steplike transition of width $\delta \phi$ and still enhance the linearized modes, this enhancement would cause them to be strongly coupled near the end of the transition where the increase in $\epsilon$ corresponds to a decrease in $\Delta N$.  
While this does not necessarily prevent the enhancement from occurring, a full calculation requires  simulations. 
By separating the transitions, our concrete model avoids this difficulty with the sharper transition that occurs before the enhancement:
$\Delta N_1 \approx \delta \phi/\sqrt{2\epsilon_1}$ whereas
the field widths associated with the other transitions  are actually much larger due to the larger $\epsilon$.

\emph{Conclusion.}---
 In this letter, we have shown that a large enhancement of perturbations results when the inflaton crosses a downward step in its potential in less than an e-fold, which counter-intuitively allows a sizable amount of PBHs to form in a model wherein the inflaton always possesses a velocity higher than its value at the horizon exit of CMB scales.
 The enhancement can be interpreted as particle production due to the non-adiabatic transition 
 whose curvature fluctuations are then adiabatically enhanced to large values as the inflaton loses the extra kinetic energy from the step due to Hubble friction.

Finally, we mention that, depending on the height and the location of the downward step, our enhancement mechanism can generate seeds not only for PBHs with a variety of masses, but also for ultra-compact minihalos~\cite{Ricotti:2009bs,Scott:2009tu,Bringmann:2011ut,Emami:2017fiy}. Additionally, the enhancement can be probed (constrained or discovered) by a range of complementary observables, such as the gravitational waves induced by the scalar perturbations~\cite{tomita1967non,Matarrese:1993zf,Matarrese:1997ay,Ananda:2006af,Baumann:2007zm,Saito:2008jc,Saito:2009jt,Kohri:2018awv,Inomata:2018epa}, and CMB spectral distortions~\cite{Fixsen:1996nj,Chluba:2012we,Kohri:2014lza}.
Future observations of PBHs and these varied observable probes will enable us to probe this characteristic feature in the inflaton potential.

\emph{Acknowledgments.}---
The authors thank Thomas Crawford, Jose Maria Ezquiaga, Hayato Motohashi, Samuel Passaglia, and David Zegeye for helpful comments.
The authors were supported by the Kavli Institute for Cosmological Physics at the University of Chicago through an endowment from the Kavli Foundation and its founder Fred Kavli.
E.M. and W.H. ware supported by U.S. Dept. of Energy contract DE-FG02-13ER41958.  W.H. was additionally supported by  the Simons Foundation.

%%%%%%%%%%%%%%%%%%%%%%%%%%%%%%%%%%%
%%%%%%%%%%%%%%%%%%%%%%%%%%%%%%%%%%%
%%%%%%%%%%%%%%%%%%%%%%%%%%%%%%%%%%%
\small
\bibliographystyle{apsrev4-1}
\bibliography{draft_pbh_arise}

\end{document}